\documentstyle[preprint,tighten,aps,epsf,psfig]{revtex}

\newcommand{\ba}{\begin{eqnarray}}
\newcommand{\ea}{\end{eqnarray}}

\newcommand{\fr}[2]{\frac{#1}{#2}}
\newcommand{\non}{\nonumber}

\newcommand{\dr}{\partial_r}

\newcommand{\lb}{\left(}
\newcommand{\rb}{\right)}
\def\vec#1{{\mbox{\boldmath$#1$}}}
\newcommand{\r}{\mbox{$\vec{r}$}}

\newcommand{\be}{\begin{equation}}
\newcommand{\ee}{\end{equation}}

\newcommand{\im}{\mbox{Im}}

\begin{document}

\preprint{BudkerINP-98-7, TTP98-10, hep-ph/9802379}

\title{Top quark production at threshold with ${\cal O}(\alpha_s^2)$
accuracy}

\author{Kirill Melnikov\thanks{
e-mail:  melnikov@particle.physik.uni-karlsruhe.de}}
\address{Institut f\"{u}r Theoretische Teilchenphysik,\\
Universit\"{a}t Karlsruhe,
D--76128 Karlsruhe, Germany}
\author{ Alexander Yelkhovsky 
}
\address{ Budker Institute for Nuclear Physics\\
Novosibirsk, 630090, Russia}
\maketitle

\begin{abstract}
We calculate the next-to-next-to-leading order
correction to  the cross section for top quark pair production
in $e^+e^-$ annihilation in the threshold region, resumming all
${\cal O}\left [ (\alpha _s/\beta )^n
\times (\alpha _s^2,\beta ^2,\alpha _s \beta ) \right ]$
terms of perturbation series.
We find that the magnitude of the NNLO correction is comparable
to the size of the NLO corrections.
\vspace{0.5cm}

\end{abstract}

\section {Introduction}

The cross section of  hadron production  in $e^+e^-$
annihilation belongs to the best known quantities in high
energy physics.  Far away from
quark thresholds the cross section is well approximated by
the results
obtained in perturbative QCD (for a  review see \cite{ChKK}).
The situation is
not so clear at quark thresholds which, however,
are known to be of importance
for a number of physical applications.

Among such applications, a special place is occupied
by a threshold production of $t\bar t$ pairs at the Next Linear Collider.
It was suggested in Ref. \cite {FKh}, that the large width of the top quark
provides a natural cutoff for long--distance effects
and, therefore,  reliable predictions  for the $t\bar t$
threshold production cross section are possible
in  perturbative QCD.  Since then, the threshold
production cross section of $t\bar t$ was studied in great
detail \cite{PS,topst1,topst2,topst3,MeYak}.
The commonly accepted conclusion \cite{topst3,topphys}
is that one can perform precision studies of various
quantities of direct physical interest (top mass, top width, strong
coupling constant, etc.), once accurate measurements in the threshold
region are conducted. However, all these studies were performed using
predictions for the top threshold production cross section valid
up to ${\cal O}(\alpha _s)$  and, therefore, suffered
from the ignorance of higher order QCD effects.

It is worth emphasizing, that calculation of radiative  corrections
to the threshold cross section differs from standard perturbative
calculations, which are done for higher energies.
The difference is because of the fact, that,
close to the threshold, the conventional perturbation theory
breaks down \cite{SS}. The physical origin of this phenomena
is known from quantum mechanics:
considering Coulomb potential as a perturbation, one gets
series in $\alpha/\beta$, where $\alpha$ is the strength
of the potential and $\beta$ is the particle velocity.
When the velocity is small, this ratio becomes large and
meaningful predictions can be only achieved once the series is resummed.
It was demonstrated in \cite {SS}, that if such resummation
is performed, the threshold  cross section
becomes proportional to the square of the Coulomb wave function
at the origin. In Ref. \cite {FKh} this result was generalized
to the situation, when the produced particles are unstable.
It was concluded, that the cross section for the
quark pair production is proportional to the imaginary
part of the non--relativistic
Green function of the $\bar Q Q$
system, evaluated for complex  energies.

Since then, it was also realized, that the
${\cal O}(\alpha _s )$ corrections can be easily incorporated,
because contributions of soft and
hard scales completely factorize with this accuracy.
The absence of this factorization
property, as well as the technical difficulties
with explicit higher order calculations, were the stumbling blocks
in  achieving the ${\cal O}(\alpha _s ^2)$ accuracy.
It is remarkable, that new results,
obtained in the last several years,
permit a relatively easy determination of these
corrections.

In what follows, we  present the calculation of the threshold
cross section for the $t\bar t$ pair production which  is
valid with ${\cal O}(\alpha _s^2, \alpha _s \beta, \beta ^2)$
accuracy.

\section {The framework of the calculation}
We first discuss a framework of our calculations and introduce
all relevant notations. The threshold region is characterized
by a small value of the quark velocity $\beta$:
\be
\beta =\sqrt{1-\frac {4m^2}{s}} \ll 1.
\ee
To order  ${\cal O}(\alpha _s^2, \alpha _s \beta, \beta ^2)$,
dynamics of  slowly moving
quark--antiquark pair is governed by a non--relativistic
Hamiltonian\footnote{ One can describe the $\bar Q Q$ system by means of the
non--relativistic quantum mechanics to this order
since  the radiation of real gluons shows up only at
${\cal O}(\beta ^3 )$ order.  }:
\begin{eqnarray}
 H &=& H_0 +V_1(r) + U(\vec{p},\vec{r}), \label{Ham}
\\
 H_0 &=& \frac {\vec{p}^2}{m} - \frac { C_F a_s }{r}, \label{H0}
\\
 V_1(r) &=& - \frac {C_F a_s^2}{4\pi r} \left\{
2\beta_0 \ln(\mu'r) + a_1  \right.
\nonumber
\\
& &
\left. + \frac{ a_s }{4\pi}
\left[ \beta _0^2 \left ( 4\ln^2(\mu'r)+\frac {\pi^2}{3} \right )
       + 2 (\beta _1 +2\beta_0a_1) \ln(\mu'r) + a_2              \right ]
       \right\},
\label{V1}
\\
 U(\vec{p},\vec{r}) &=& -\frac {\vec{p}^4}{4m^3}
 + \frac {\pi C_F a_s }{m^2} \delta^{(3)} ( \r )
-\frac {C_F a_s}{2m^2r}
\left (\vec{p}^2 + \frac {\vec{r} (\vec{r}\vec{p}) \vec{p}}{r^2} \right )
\nonumber \\
&&+\frac {3C_F a_s}{2m^2r^3} \vec{S} \vec{L}
-\frac {C_F a_s }{2m^2} \left ( \frac {\vec{S}^2}{r^3}-
3\frac { \lb \vec{S} \vec{r} \rb^2}{r^5} - \frac {4\pi}{3}
(2 \vec{S}^2 -3) \delta^{(3)}
(\vec{r} ) \right )-\frac {C_AC_F a_s^2}{2mr^2}.
\label {Breit}
\end{eqnarray}
In the above equations, the strong coupling constant  is
evaluated at the scale $\mu$:
\be
a_s \equiv \alpha_s (\mu).
\ee
The scale $\mu '$ equals to $\mu e^{\gamma}$, $\gamma$ is the Euler
constant.

The operator $U(\vec{p},\vec{r})$ is the QCD generalization of the
standard Breit potential \cite {LL}. The last term in Eq.(\ref{Breit})
is the non--Abelian contribution, originating from a
correction to the Coulomb gluon exchange, caused  by a magnetic gluon
\cite {NABreit}.
The potential $V_1(r)$ represents a deviation  of the static
QCD potential from the Coulomb one. It was
calculated to order $\alpha _s ^2$
in \cite {Fischler} and to order $\alpha _s ^3$
in \cite {Peter}.
The coefficients there read explicitly:
\begin{eqnarray}
\beta _0 &=& \frac {11}{3}C_A-\frac {4}{3}N_LT_R,
\nonumber \\
\beta _1 &=& \frac {34}{3} C_A^2 - \frac {20}{3} C_AT_RN_L - 4C_F T_R N_L,
\nonumber \\
a_1 &=& \frac {31}{9} C_A - \frac {20}{9} T_R N_L,
\nonumber \\
a_2 &=& \left (\frac {4343}{162} + 6\pi^2 - \frac {\pi^4}{4}+\frac
{22}{3} \zeta_3 \right )C_A^2 -
\nonumber \\
&&
\left (\frac {1798}{81}+\frac {56}{3}\zeta_3
\right ) C_A T_R N_L
-\left (\frac {55}{3} - 16\zeta_3 \right )C_FT_RN_L + \left (\frac
{20}{9} T_R N_L \right )^2.
\end{eqnarray}

For the $SU(3)$ color group, the color factors are
$C_A = 3, C_F = 4/3, T_R = 1/2$. $N_L = 5$ is the number of quarks
whose masses have been neglected.

Given the Hamiltonian $H$, one can find the  Green function
for the Schr\"odinger equation:
\be
(H-E-i\delta)G(E;\r,\r') = \delta^{(3)}(\r -\r').
\ee

Once the Green function is found, the
cross section of the non--relativistic $Q\bar Q$ pair
production in $e^+e^-$
annihilation\footnote{
In what follows, we consider only photon mediated process
and do not take into account the $Z$--boson exchange.
The axial-vector coupling of the
$Z$--boson contributes ${\cal O}(\beta ^2)$ relative correction
to the threshold cross section. The vector $Z e^+ e^-$
coupling is also suppressed, but can be taken into account
in the same way as the photon contribution, which we treat
in this paper.}
is obtained as:
\be
\sigma (s)  = \frac {4\pi\alpha ^2}{3s} R(s),
\ee
where
\be
R(s) = \lim _{r \to 0} {\mbox {Im}} \left [ N_c e_Q^2 \frac {24\pi}{s}
\left (1-\frac {\vec{p}^2}{3m^2} \right )
G(E;\r,0) \right ],\qquad E=\sqrt{s}-2m.
\label {crsect}
\ee
In Eq.(\ref {crsect}),
we have included the ${\cal O}(\beta^2)$ correction
originating from the expansion of the vector current which
produces and annihilates a heavy $Q\bar Q$ pair in the triplet
$S$--state. The quantity $R(s)$ will be the central object for  further
discussion.

If a calculation of $R$
will be attempted, one will find, that
the Green function at the origin does not exist because
there are terms in the Hamiltonian $H$, which behave
as $1/r^{n},\;n \geq 2$,  for small values of $r$.
The difficulty originates from the fact, that
the region $r \to 0$ is not properly treated in the Hamiltonian.
Indeed, small values of $r$ correspond to a region
in the momentum space, where
a typical momentum transfer
between $\bar Q$ and $Q$, is of the order of the quark
masses and therefore  quarks
cannot be considered as non--relativistic.
For this reason, the use of the Hamiltonian $H$ in actual
calculations leads to the divergencies, which appear for
$r \to 0$.

The way to circumvent this difficulty is as follows. In order to
perform a calculation, one introduces
a cutoff $\lambda $,
such that $\alpha _s m \ll \lambda \ll m$. The momenta region where
$k \ll \lambda$ is a non--relativistic region and can be described
using a Hamiltonian $H$. The momenta region with  $k \gg \lambda$
is a relativistic one and the calculation in this region should
be performed using  the  rules of  quantum field theory.
We note that this is rather
standard procedure for calculations, related to bound state
problems. It is also well known
that its practical realization often requires substantial effort.

However, there is  a possibility to use the result of the
non--relativistic calculation with the cutoff in the following way:
one takes the limit of the obtained result,
considering kinematic region where
$\alpha _s \ll \beta \ll 1$. In this particular region,
the non--relativistic results
are still valid; on the
other hand as long as $\alpha_s/\beta \ll 1$, the resummation
of the Coulomb effects is not necessary.
Therefore, in this particular region, one can
calculate the corrections applying the standard rules of the quantum field
theory.
In the framework of QCD, such calculations
have been performed recently  in Ref. \cite{CzM}.
One therefore can match the result of the non--relativistic
calculation with the cutoff, directly to
the result presented in \cite{CzM} and in this way
completely eliminate the cutoff dependence.
This procedure was suggested in \cite{Hoang} and we will call it a
direct matching procedure, in accordance with  that reference.

In what follows, we pursue this program in QCD.
We confine ourselves to a strictly perturbative approach
and we do not attempt any discussion of non--perturbative
effects. In order to accommodate
the phenomenologically relevant
case of the unstable top quark, we will consider the total
energy $E$ as the complex variable
$E \to E + i\Gamma _t$, in the spirit
of Ref. \cite {FKh}.

We have to mention at this point that a consistent implementation
of the effects related to the finite width of the quark is not
attempted in this paper. To ${\cal O}(\alpha _s)$ order
such effects were studied in \cite {MeYak}. A reliable
treatment of these effects to ${\cal O}(\alpha _s ^2)$
is not available at the moment.

\section{Corrections to the Green function at the origin}

Let us first consider the Hamiltonian $H_0$,  Eq.(\ref{H0}), as the
leading order Hamiltonian. The corresponding Green function
will be denoted by $G_C(\r,\r')$.

We first discuss a correction to the Green function  $G_C(\r,\r')$
at the origin, caused by the operator $U(\vec{p},\vec{r})$
(see Eq.(\ref {Breit})).
The first order correction is:
\be
-\int {\rm d}^3 r\; G_C(\r,0)\;U(\vec{p},\vec{r})
\;G_C(\r,0).
\label {pert}
\ee

As long as we are interested in the $Q\bar Q$ pairs, produced
in the triplet $S$--states, only a corresponding
projection of the operator $U(\vec{p},\vec{r})$ should be considered.
Substituting $\vec{S}^2=2 $ and $\vec{L}=0$ into Eq.(\ref {Breit}),
we get:
\be
U(\vec{p},\vec{r}) =
-\frac {\vec{p}^4}{4m^3} + \frac {11\pi C_F a_s}{3m^2} \delta^{(3)} (\r )
-\frac {C_F a_s}{2 m^2}\left \{\frac {1}{r},\vec{p}^2 \right \}
-\frac {C_AC_F a_s^2}{2mr^2}.
\label {Br}
\ee

At this stage, it is advantageous to express this operator in terms of
the zeroth order Hamiltonian $H_0$ in order to apply the equation
of motion for the Green function  $G_C(\r,0)$:
\be
(H_0-E)G_C(\r,0) = \delta^{(3)}(\r).
\label {eqmot}
\ee

This is most easily done by using the commutation relations:
\ba
 \left [ H_0, ip_r \right ] &=& \fr{4\pi\delta ^{(3)}(\r)}{m}+
 \fr{2\vec{L}^2}{mr^3}-
 \fr{C_F a_s}{r^2},
\\
\left\{H_0,\fr{1}{r}\right\}&=&\fr{2}{r}H_0+
\fr{4\pi\delta ^{(3)} (\r)}{m}+\fr{2}{mr^2}\dr,
\ea
where $p_r = -i (\dr + 1/r )$ is the radial momentum operator.
Using these  relations, one finds that the operator $U(\vec{p},\r)$
from Eq.(\ref{Br}) can be presented in the form:
\be
\label{Umod}
U(\vec{p},\r) = - \fr{ H_0^2 }{4m}
              - \fr{ 3C_F a_s }{4m} \left\{H_0,\fr{1}{r} \right\}
              + \fr{ 11 C_F a_s }{12m} \left [ H_0, ip_r \right ]
              - \fr{ \lb 2C_F + 3C_A \rb C_F a_s^2 }{ 6mr^2 }.
\ee

Let us consider the first three terms in  Eq.(\ref {Umod}).
Inserting them into Eq.(\ref {pert}) and using the equation of motion
for the Green function (\ref {eqmot}),
we find:
\ba
&&- \int {\rm d}^3 r G_C(\r',\r)
  \lb -\fr {H_0^2}{4m}- \fr{ 3C_F a_s }{4m} \left\{H_0,\fr{1}{r} \right\}
              + \fr{ 11 C_F a_s }{12m} [ H_0, ip_r] \rb G_C(\r,\r'')
\nonumber \\
&& = \left [ \fr {\vec{p}^2}{2m^2} + \fr{ C_F a_s }{mr'}
             + 2ip_r \right ] G_C(\r',\r'')
    + \int {\rm d}^3 r G_C(\r',\r) \left \{ \fr {E^2}{4m} +
\fr{ 3C_F a_s E }{2mr} \right \} G_C(\r,\r'').
\label{cc}
\ea

We note, that all terms in the above equation, which  do not
contribute to the imaginary part of the Green function have been
omitted.

Leaving aside the ``surface'' terms in Eq.(\ref{cc}), one sees
that the discussed perturbation can be absorbed into
the zeroth order equation (\ref{eqmot}). Therefore,
the corresponding correction to $G_C(\r,\r')$ can be taken into account
to all orders by finding an exact Green function
for the Schr\"odinger equation
\be\label{GFeq}
\left ({\cal H} - {\cal E} \right ) G(\r,\r') = \delta^{(3)} (\r-\r'),
\ee
with the modified Hamiltonian
\be
{\cal H} = \frac {\vec{p}^2}{m} +V(r),\qquad
V(r) =
- \frac {C_F a_s }{r}\left (1+\frac {3E}{2m}
\right )
+V_1(r) -\frac { (C_F a_s )^2}{2mr^2}
\left (\frac {2}{3} + \frac {C_A}{C_F} \right ),
\label {Vr}
\ee
and the modified eigenvalue
\be\label{Emod}
{\cal E} = E + \frac {E^2}{4m}.
\ee

It is clear, that the solution of the above equation
will deliver $G(\r, \r')$, which
is definitely valid to NNLO accuracy.
Moreover, such a solution provides a resummation of some
second order corrections.

But there is more important reason to believe that such a treatment
is more appropriate for the subthreshold
($\sqrt {s} < 2m$)
energy region, than the
first order time--independent perturbation theory.
Let us consider the energy region below the threshold. For stable
quarks, one would observe an appearance of narrow resonances
in this region. It is well known, that for the realistic
value of the top quark width ($\Gamma _t \sim 1.5\; {\rm GeV}$),
the resonances are smeared. Still, the excitation curve exhibits
a maximum close to the position of the lowest lying resonance.

For the purpose of discussion, we write an expression for an
exact Green function:
\ba
 G(E+i\Gamma;0,0)
=
      \sum_n \fr{ \psi_n^2 (0) }{ E_n - E - i\Gamma }
              + \int_0^{\infty} \fr{ dk }{ 2\pi }
                     \fr{ \psi_k^2 (0) }{ E_k - E - i\Gamma },
\label {rGF}
\ea
where $\psi _{n,k}$ and $E_{n,k}$ are exact eigenfunctions and
eigenvalues, which correspond to a Hamiltonian $H$.
When we calculate this Green function using time--independent
perturbation theory, we expand both numerators and denominators
of the above equation in power series. One readily realizes,
that this procedure is not so harmless for  energy
denominators, especially when the energy $E$
is close to the position of the resonance.
On the other hand, if the Green function is obtained
as the solution of the Schr\"odinger equation, one gets
the result directly in the form of Eq.(\ref {rGF}).
It is mainly
for this reason, that we think it is more safe to solve
the Schr\"odinger equation exactly,
than to perform the first order perturbation theory calculations.

It is not so straightforward, however, to obtain a
numerical solution for such a problem, since the perturbation $1/r^2$
(cf. Eq.(\ref {Vr}))
is too singular at the origin.
As was already indicated above, the proper approach is to introduce
a cutoff $r_0$ and to extract all terms which have a non--analytic
dependence with respect to $r_0$. On the other hand, all terms
which have extra powers of $r_0$, so that
the limit $r_0 \to 0$
can be taken,  will be set to zero.
Later on, the non--analytic $r_0$--dependent
terms will be determined by a matching of the result
of such  calculation to
its perturbative counterpart \cite{CzM}.
In the next section we will show, how this procedure
can be implemented for the numerical solution of the Schr\"odinger
equation (\ref{GFeq}).

The ``surface'' terms
from Eq.(\ref {cc}) will be discussed later. We note here, that these
terms are linear in $G_C(\r, \r')$ and, therefore,
contain at most the first--order poles in energy. 
Hence they do not contribute to the shift of the energy levels and
there is no need to account for them beyond the first order.

\section{Numerical solution of the Schr\"odinger equation}
\label {four}

In this section, we demonstrate, how  numerical solution of the
Schr\"odinger equation with the potential $V(r)$
(cf. Eqs.(\ref{GFeq}) and (\ref {Vr})) can be constructed.
The Schr\"odinger
equation for the $S$--wave Green function
is written as:
\be
\left ( \frac {-1}{m} \left [ \frac {{\rm d}^2}{{\rm d}r^2}
+ \frac {2}{r} \frac {{\rm d}}{{\rm d}r} \right ]
+V(r) - {\cal E} \right )G(r,r') = \frac {1}{4\pi r^2} \delta (r-r').
\ee

It is convenient to define a new function $g(r,r')=rr'G(r,r')$,
so that the Schr\"odinger equation simplifies:
\be\label{eqforg}
\left (  \frac {{\rm d}^2}{{\rm d}r^2}
+ m \left [{\cal E} -V(r) \right ] \right )g(r,r') = - \frac {m}{4\pi }
\delta (r-r').
\ee

According to the standard rules, the Green function can be written as:
\be
g(r,r') = A \left [ g_<(r)g_>(r')\theta (r'-r) +
g_<(r')g_>(r) \theta (r-r') \right ]
\ee
where $g_{<,>}(r)$ are two independent
solutions of the homogeneous Schr\"odinger equation which
satisfy proper boundary conditions at $r=0$ and $r=\infty$,
respectively. Also, the constant $A$ is defined by the
jump of the derivative of the Green function at $r=r'$.

The solutions $g_<(r)$ and
$g_>(r)$
are constructed using  two other
independent solutions of the Schr\"odinger equation,
$g_{\pm}(r)$,  with a prescribed behavior
at the origin \cite {PS}.
However,
since the potential $V(r)$ in the Schr\"odinger equation is as singular
as $1/r^2$, one would have problems with setting the standard \cite{PS}
boundary conditions for $g_{\pm}$ at $r=0$.
To overcome this
difficulty,  we extract the leading asymptotic of the
functions $g_\pm(r)$ for small values of $r$
\be
g_{\pm}(r) = (mC_F a_s r)^{d_{\pm}} f_{\pm}(r),
\label {gtof}
\ee
where
\be
d_{\pm} = \frac {1}{2} \left ( 1 \pm \sqrt{1- 4\kappa} \right ),\qquad
\kappa = \frac {(C_F a_s )^2}{2}
\left (\frac {2}{3} +\frac {C_A}{C_F} \right ).
\label {dpm}
\ee
Substituting $g_\pm (r)$ to the Schr\"odinger equation, one obtains
an equation for the function $f_\pm (r)$, which is now free
of the $1/r^2$ term:
\be
\left \{ \frac {1}{m} \left [ \frac {{\rm d}^2}{{\rm d}r^2}
+ \frac {2d_\pm }{r} \frac {{\rm d}}{{\rm d}r} \right ]
+ \left [ {\cal E} + \frac {C_Fa_s}{r} \left (
1+c_1\ln(\mu 'r)^2 + c_2 \ln(\mu ' r) + c_3 \right ) \right ]
\right \} f_\pm (r) =0.
\label {Eqf}
\ee
The coefficients $c_{1-3}$ can be easily  obtained
using Eq.(\ref {Vr}) and Eq.(\ref{V1}).
One derives then the asymptotics of the function $f_{\pm}(r)$
for $r \to 0$:
\begin{eqnarray}
f (r) &\sim& 1 - m C_F a_s r A(r) + {\cal O}(r^2)
,\qquad A (r) =
 h_1 \ln^2( \mu ' r ) + h_2 \ln (\mu ' r) + h_3 ,
\label{A} \\
h_1 &=& \fr{ c_1 }{ 2d },
\label {h1} \\
h_2 &=& \fr{ 1 }{ 2d } \left[ c_2 - c_1 \lb 2 + \fr{ 1 }{ d } \rb
\right],
\label {h2} \\
h_3 &=& \fr{ 1 }{ 2d } \left[ 1 + c_3 - c_2 \lb 1 + \fr{ 1 }{ 2d } \rb
+ c_1 \lb 2 + \fr{ 1 }{ d } + \fr{ 1 }{ 2d^2 } \rb\right].
\label {h3}
\end{eqnarray}
In the above equations,  $f$ and $d$ stand for $f_{\pm}$ and $d_{\pm}$,
respectively.

There could be some doubts about the validity of the boundary conditions,
as derived from above equations, since
$d_-$ is of
order
$\alpha _s ^2$
and therefore the second power
of $\alpha _s $ appears in the
{\em denominator}.
Without going into
explanations at this point, we mention that upon careful
inspection the above boundary conditions
appear to be absolutely legitimate. Later, we will present more detailed
arguments in favor of such conclusion.

The Green function is constructed  as follows.
The solution $g_<(r)$ is identified with  $g_{+}(r)$. The solution
$g_>(r)$ is constructed from the solutions $g_{\pm}$ in such a way,
that the boundary condition at the infinity is satisfied:
\be
g_{>}(r) = g_{-}(r) + B\;g_{+}(r),\qquad g_{>}(r) \to 0,\qquad {\rm for}
\;r \to \infty.
\ee
Therefore, one finds:
\be
B = - \lim _{r \to \infty} \left [ \frac {g_-(r)}{g_+(r)} \right ].
\label {B}
\ee

If the potential $V(r)$ is {\rm real}, 
the imaginary part of the Green function
is proportional to the imaginary part of the coefficient $B$.
However, this is not the case for the present problem:
the Coulomb
part of the potential $V(r)$ is energy dependent and we consider
the energy to be a complex variable. Therefore, the formula
for the Green function at the origin should be modified.
The modification is however simple. It is obtained
in a straightforward way from the available asymptotics of
the functions $\psi _{\pm} (r)$. The result reads
\be
\im\; G(r_0,r_0) = - \frac {m^3(C_F a_s)^2}{4\pi W } \im \left \{
 ( m C_F a_s r_0 )^{2d_+ -2 }\; B
-  A_+ (r_0) - A_- (r_0)  \right \},
\label {GF}
\ee
where $W = - mC_F a_s (d_+ - d_-)$ is the Wronskian
of  two independent solutions of the Schr\"odinger equation.

Let us comment on the role of the second term
in the above equation (\ref {GF}). If the potential $V(r)$ in the
Schr\"odinger equation  were real, the $A_{\pm}(r)$ would be real
as well.
Taking the imaginary part, one then completely removes the
contribution of the second term in Eq.(\ref {GF}).
In our case $E$ is a complex variable, and the coefficient $h_3$
has a non--zero imaginary part.
Moreover, this imaginary part
is formally of the order of $\Gamma _t/(md_-)$,
which should be considered as a contribution of order unity.

However, the limit $\kappa \to 0$
should exist for
the Green function constructed as above.  Therefore, the purpose
of the last term in Eq.(\ref {GF}) is to cancel the
$1/d_-$ singularities of the function $B$, obtained with the boundary
conditions presented in Eqs.(\ref{h1}--\ref{h3}).
We note in this respect, that, by switching
off the logarithm--dependent perturbations in the expression for
$V(r)$, one obtains an exactly solvable Hamiltonian, so that  the statements
made above can be easily verified. The corresponding
discussion can be found in Appendix.

Eq.(\ref {GF}) then explicitly demonstrates,
that the non--analytic dependence on the cutoff $r_0$ is indeed
extracted and all power corrections with respect to the cutoff are
neglected. This precisely corresponds to the desired form of the
Green function of the Schr\"odinger equation.  The residual dependence
on $r_0$ is removed using the direct matching procedure as
described in the next section.

\section{Matching and final result for $R$}

In Eq.(\ref{GF}) we have taken into account all corrections to the
ratio $R$ which are due to relativistic effects in the quark--antiquark
interaction and which can thus be called the dynamic ones.
However, we still have to consider kinematic corrections
that are: i) ${\cal O}(\beta^2)$
correction to $s$; ii) ${\cal O}(\vec{p}^2/m^2)$ correction to the quark
current (cf. Eq.(\ref{crsect}))  and, finally, iii) the "surface''
terms from Eq.(\ref{cc}).

Using equation of motion for the Green function  (\ref{GFeq}), we
obtain the result for $R$ at NNLO,
with both types of corrections included\footnote{We
note that, strictly speaking,  the ``surface'' terms in Eq. (\ref{cc})
were derived for the Coulomb Green function, but we substitute
the ``exact'' Green function instead of the Coulomb one in
our final formulas.}:
\ba\label{NNLO}
R_{\mbox{\scriptsize{NNLO}}} &=& \fr{3}{2} N_c e_Q^2 C_F a_s
     \lb 1+ C_1 C_F \left ( \fr{ a_h }{\pi} \right ) +
     C_2 C_F \lb \fr{ a_h }{ \pi } \rb^2 \rb
     \non \\
     && \times \fr{ 1 }{ \sqrt{1 - 4\kappa} }
        \mbox{Im} \left \{ \lb 1 - \fr{ 5\beta^2 }{ 6 } \rb
        \left [ (m C_F a_s r_0)^{\sqrt{1 - 4\kappa}-1} B
         -  A_+ (r_0) -  A_- (r_0)  \right ] \right \}.
\label{FR}
\ea
Here we have factored out all energy--independent corrections. They are
parametrized by the
constants $C_1$ and $C_2$, which are divergent
\footnote{The strongest divergence in these factors is $1/r_0$,
so it not obvious  {\it a priori}, that the functional
form of Eq.(\ref {GF}) can be preserved in Eq.(\ref{FR}).
Upon careful analysis, this turns out to be possible
to NNLO, however.}
in the limit
$r_0 \rightarrow 0$. For this reason
we use $a_h = \alpha_s(m)$ as the expansion
parameter for these ``hard'' corrections.

To get rid of the $r_0$--dependence we use the direct matching
procedure, suggested in Ref. \cite{Hoang}.
For this we consider $\sqrt {s} > 2m$, set the width of the top
quark $\Gamma _t$  to zero\footnote{Note, that in this case
the functions $A_\pm(r_0)$  drop out from Eq.(\ref {NNLO}).}
 and
equate our result (\ref{NNLO}) to its perturbative counterpart
\cite{CzM} in the kinematic region $\alpha_s \ll \beta \ll 1$,
where both are supposed to be valid.
We also set $\mu = m$, so that
$a_s$ coincides with $ a_h $.

Let us note, that the direct matching procedure fixes
the linear combination of $C_2$ and $\ln (mr_0)$
\be
C_2 C_F \lb \fr{\alpha _s}{\pi} \rb^2 - 2\kappa \ln(mr_0).
\ee
If we were working strictly
to NNLO, this last combination would be the only thing we need
for the final result. However, because of the large 
difference in scales, which
govern relativistic and non--relativistic physics,
we would like to write Eq.(\ref {FR}) in a
factorized form and include an exact dependence on $r_0$ into
the non--relativistic Green function. For this reason, we
have to  set a factorization scale. We  do this by choosing
$r_0$ in such a way, that the correction to the Coulomb Green
function due to the $1/r^2$ perturbation
in the region $\alpha _s \ll \beta \ll 1$ is given
by $\log(\beta)$, without additional constants
(see Appendix for more details). Any other choice
of $r_0$ would correspond to other (also legitimate)
value of the factorization  scale.

A factorized form (\ref {FR}) of our final result 
makes sense only if a dependence
on a choice of the factorization scale is weak. We have
checked that changing the value of the cutoff between
$r_0/2$ and $2r_0$ for $r_0$ given by Eq.(\ref{r0}),
we obtain small ($\sim 1-3\%$) variation of the resulting values of
$R$.

Therefore, accoring to our choice, we fix the value of the cutoff
\be
r_0 = \frac{e^{2-\gamma }}{2m},
\label {r0}
\ee
and obtain finally:
\be
C_1 = -4; \qquad
C_2 = C_F C_2^{A} + C_A C_2^{NA}
+T_R N_L C_2^{L} + T_R N_H C_2^{H},
\label {Co}
\ee
where
\begin{eqnarray}
C_2^{A}&=&\frac {39}{4} -\zeta _3 +\pi ^2 \left (\frac
{4}{3}\ln2-\frac {35}{18} \right );
\nonumber \\
C_2^{NA}&=& -\frac {151}{36}-\frac{13}{2} \zeta _3
+\pi^2 \left (\frac {179}{72} -\frac {8}{3}\ln2 \right );
\nonumber \\
C_2^{H} &=&\frac {44}{9} - \frac {4}{9}\pi^2;
\nonumber \\
C_2^{L} &=& \frac {11}{9}.
\label {hard}
\end{eqnarray}

Eq.(\ref {NNLO}) with definitions provided by
Eqs.(\ref {r0} --\ref{hard}) is our final result
for the top quark threshold cross section with the NNLO accuracy.

For numerical purposes, we have chosen
$m = 175~{\rm GeV}$ and
$\Gamma _t = 1.43~{\rm GeV}$. As an input value for the
strong coupling constant we used $\alpha _s (M_Z) = 0.118$.
Fig.1 provides our final results for 
$R_{\mbox{\scriptsize{NNLO}}}$ as a function of $\sqrt{s}-2m$
in comparison with LO and NLO results, for three
values of the soft scale $\mu  = 50,~75,~100~{\rm GeV}$.
One can see that the NNLO corrections are as large as the
NLO ones. 

There is also a moderate scale dependence of the
NNLO corrections in the vicinity of the resonance peak.
The position of the resonance peak appears to be sensitive
to the variations in the scale $\mu$ on the level
$\sim 100\; {\rm MeV}$. We note in this respect, that the
shift of the ground--state energy due to the Breit perturbation
is well known (see Ref. \cite {LL} and Appendix) and its expected
variation with $\mu$ is close to this value.

\section {Conclusions}
We have presented a calculation of the next-to-next-to-leading order
corrections to the threshold cross section of the top quark pair
production in QCD, summing  all
${\cal O}\left [ (\alpha _s/\beta )^n
\times (\alpha _s^2,\beta ^2,\alpha _s \beta ) \right ]$
terms of the perturbation series. We have found, that the NNLO
effects are quite sizable.

We have also discussed how the numerical solution of the Schr\"odinger
equation with a singular potential can be constructed.

Let us comment on the large size of the NNLO corrections.
We have checked that taken separately, both the Breit perturbation and 
${\cal O}(\alpha _s ^3)$
terms from $V_1(r)$, provide comparable
contributions of the same sign. When we take them
into account  simultaneously
in the Schr\"odinger equation, the  NNLO contribution
gets enhanced by roughly a factor of two, in the vicinity of the resonance.

When we were writing this paper, we received the preprint
\cite {HT}, where the same problem was studied. Our qualitative
conclusion about the size of the NNLO corrections agrees with the
conclusion reached in \cite {HT}.

\section*{ Acknowledgments}
We are grateful to Y.Sumino for useful comments.
K.M. would like to thank A. Czarnecki and R. Harlander
for advices concerning numerical calculations.
The research of K. M. was supported in part by BMBF
under grant number BMBF-057KA92P, and by Graduiertenkolleg
``Teilchenphysik'' at the University of Karlsruhe.

\section*{Appendix}

In this Appendix, we discuss the construction
of the Green function $G({\cal E}; r_0,r_0)$ in the exactly solvable model,
which is described by the Hamiltonian
\be\label{Hmodel}
H  = \frac {\vec{p}^2}{m} +V(r),\qquad
V(r) =
- \frac {C_F a_s }{r} ( 1 + c )
-\frac {\kappa }{mr^2},
\ee
where ${\cal E}$ and $c$ are complex parameters.

The Green function is derived following our discussion in
Sect. \ref {four}. It is convenient to introduce a new
variable $z = C_F a _s m  r$ for further discussion.
The solution of Eq.(\ref{Eqf}), which
satisfies boundary conditions given by  Eq.(\ref{A}),
can be written as
\be
f_\pm(z) = e^{\frac {iz}{2\nu}} F\left (
d_\pm - i\nu ( 1 + c ) ,\;2d_\pm,\;-\frac {iz}{\nu} \right ),
\ee
where $\nu = C_Fa_s /\lb 2\sqrt{ {\cal E}/m }\rb $. Using the
asymptotic form  of the confluent hypergeometric function for
$\mbox{Re}\; x = \mbox{Re}(-iz/\nu) \gg 1$,
\be
F (a,b,x) \sim \fr{ \Gamma (b) }{ \Gamma (a) } e^x x^{a-b},
\ee
one  obtains the coefficient $B$, using
Eqs.(\ref {gtof}) and (\ref {B}):
\be
B(\nu) = \fr{ i }{ \nu } (i\nu)^{2-2d_+} \fr{ \Gamma (2d_-)
         \Gamma (d_+ - i\nu( 1 + c )) }{
         \Gamma (2d_+) \Gamma (d_- - i\nu( 1 + c ))},
\ee
The functions $A_{\pm}(r)$ reduce now to the constants:
\be
A_{\pm}(r) = \fr{ 1 + c }{ 2d_{\pm} }.
\ee

We therefore arrive at the final expression for the imaginary part
of the Green function $\im G(r_0,r_0)$ for this model  
(cf. Eq.(\ref{GF})):
\be
\im G(r_0,r_0) =  \frac { m^2 C_F a_s
                   }{
                    4\pi (d_+ - d_-)}\im
                    \left\{ \left (m C_F a_s r_0 \right )^{2d_+ - 2}
                    B(\nu) - \frac{1+c}{2\kappa} \right\}.
\label {GFmodel}
\ee

Let us first demonstrate that the proper limit $\kappa \to 0$
exists for the imaginary part of the
Green function defined through Eq.(\ref {GFmodel}).
In this limit, our model reduces to
the ordinary  Coulomb problem, so that the Green
function in Eq.(\ref{GFmodel}) should give us the imaginary
part of the Coulomb Green function at the origin. To see how this
happens, we expand the Green function Eq.(\ref{GFmodel})
in power series in $\kappa $. The first term in the expansion of
$B(\nu)$ is equal to  $(1 + c)/(2\kappa)$.
This term is completely canceled by the last term in Eq.(\ref{GFmodel}).
The next term in the expansion provides the $r_0$--independent result:
$$
\lim _{\kappa \to 0} G(r_0,r_0) = \frac {m^2 C_F a_s}{4\pi}
\im H(\nu, \beta),
$$
where
\be
H(\nu,\beta) =
 \frac {i}{2\nu} - (1+c) \left[ \gamma + \ln (-i\beta)
+ \psi \left (1 - i \nu (1+c) \right ) \right ] ,\qquad
\psi (z) = \frac {{\rm d}}{{\rm d}z} \ln \Gamma (z) ,
\ee
which exactly coincides with the imaginary part of the Coulomb Green
function at the origin.

We then set $c=0$ and  expand Eq.(\ref {GFmodel}) up to
${\cal O}(\kappa)$  to obtain the correction
to the Coulomb Green function due to the $1/r^2$ perturbation.
The result can be written as:
\be
\im \left [ \delta  G(r_0,r_0) \right ]
= \frac {m^2 C_F a_s \kappa }{4\pi}
\im \left \{ 
H(\nu, \beta)^2 + \frac {\beta^2}{(C_F a_s)^2}
\right \},
\label{H2}
\ee
where the value of $r_0$ from  Eq.(\ref{r0}) has been used.
For stable quarks, the last term in Eq.(\ref{H2}) can be disregarded.
For unstable quarks, it contributes  an
${\cal O}(\Gamma _Q/m_Q)$
relative
correction
to the Green function at the origin, which is beyond the intended
accuracy and can be omitted.

Let us also emphasize one advantage of the
Green function as obtained from Eq.(\ref {GFmodel}).
Consider the stable quark case. Then, for negative
energies, the Green function should deliver the
first order poles which correspond to the appearance
of the $Q \bar Q$ bound states in the spectrum.
Eq.(\ref {GFmodel}) shows, that such poles are provided
by the singularities of the function $\Gamma(d_+-i\nu(1+c))$.
The corresponding eigenvalues of the Hamiltonian (\ref{Hmodel}) are
\be
{\cal E}_n = - \frac {m (C_F a_s)^2 (1+c)^2 }{4 (n-d_-)^2}.
\ee
Using the relations (cf. Eqs. (\ref{Vr}) and (\ref{Emod}))
\be
{\cal E} = E + \fr{E^2}{4m},  \qquad   c = \fr{3E}{2m},
\ee
as well as
\be
d_{\pm} = \frac {1}{2} \left ( 1 \pm \sqrt{1- 4\kappa} \right ),\qquad
\kappa = \frac {(C_F a_s )^2}{2}
\left (\frac {2}{3} +\frac {C_A}{C_F} \right ),
\ee
one easily finds that the energy levels are located
at
\be
E_n \approx - \fr{ m \lb C_F a_s \rb^2 }{ 4n^2 }
        +  \fr{ m \lb C_F a_s \rb^4 }{ n^3 }
           \left[ \fr{ 11 }{ 64n } - \fr{1}{6} - \fr{C_A}{ 4C_F } \right].
\ee
We note, that this is precisely what one gets, if the energy shift
due to the perturbation (\ref{Breit}) is calculated using the standard
rules of quantum mechanics.


\begin{figure}
\begin{center}
    \leavevmode
    \epsfxsize=11.cm
    \epsffile[100 250 540 530]{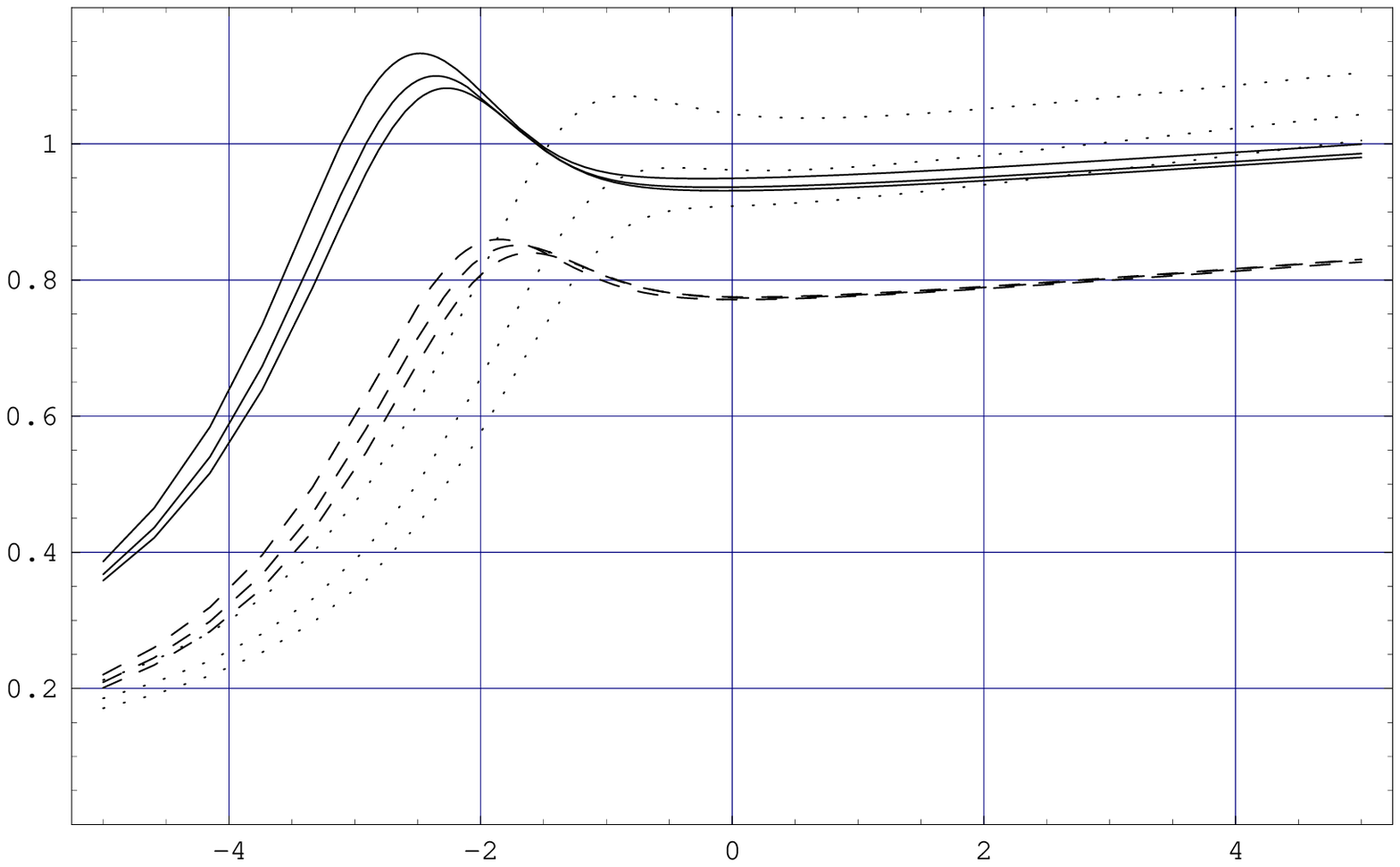}
    \hfill
    \parbox{14.cm}{\small\bf
    \caption[]{\label{figxi4}\sloppy
     {\it $R_{\rm LO}$ (dotted lines),
 $R_{\rm NLO}$ (dashed lines),
 $R_{\rm NNLO}$ (solid lines) as a function of energy
$\sqrt{s} -2m$, GeV.
In all three cases,
three curves correspond to different choices of the soft
scale $\mu  = 50\; {\rm GeV}$ (upper curves),
 $\mu  = 75\; {\rm GeV}$ and
 $\mu = 100\; {\rm GeV}$ (lower curves).
We also use  $m = 175\; {\rm GeV}$, $\Gamma _t = 1.43\; {\rm GeV}$
and $\alpha _s (M_Z) = 0.118$ as the input parameters.
 } }}
  \end{center}
\end{figure}

\end{document}